\documentclass[aps,prl,twocolumn]{revtex4-1}

\newcommand{\ket}[1]{|#1\rangle}

\newcommand{\unit}[1]{\ensuremath{\;\mathrm{#1}}}

\usepackage{bbold}
\usepackage{graphicx}
\usepackage{hyperref} 
\hypersetup{bookmarks=true,colorlinks=true,linkcolor=blue,citecolor=green}

\begin{document}


\title{An Experimental Implementation of Oblivious Transfer in the Noisy Storage Model}

\author{C. Erven$^{1,2}$}
\email[]{chris.erven@bristol.ac.uk}
\author{Nelly Huei Ying Ng$^{3}$}
\author{N. Gigov$^{1}$}
\author{R. Laflamme$^{1,4}$}
\author{S. Wehner$^{3,5}$}
\author{G. Weihs$^{1,6}$}
\affiliation{$^1$Institute for Quantum Computing and Department of Physics \& Astronomy, University of Waterloo, Waterloo, ON, N2L 3G1, Canada}
\affiliation{$^2$Centre for Quantum Photonics, H. H. Wills Physics Laboratory \& Department of Electrical and Electronic Engineering, University of Bristol, Merchant Venturers Building, Woodland Road, Bristol, BS8 1UB, UK}
\affiliation{$^3$Center for Quantum Technologies, National University of Singapore, 2 Science Drive 3, 117543, Singapore}
\affiliation{$^4$Perimeter Institute, 31 Caroline Street North, Waterloo, ON, N2L 2Y5, Canada}
\affiliation{$^5$School of Computing, National University of Singapore, 13 Computing Drive, 117417 Singapore}
\affiliation{$^6$Institut f\"ur Experimentalphysik, Universit\"at Innsbruck, Technikerstrasse 25, 6020 Innsbruck, Austria}

\date{\today}

\begin{abstract}
Cryptography's importance in our everyday lives continues to grow in our increasingly digital world. Oblivious transfer (OT) has long been a fundamental and important cryptographic primitive since it is known that general two-party cryptographic tasks can be built from this basic building block. Here we show the experimental implementation of a 1-2 random oblivious transfer (ROT) protocol by performing measurements on polarization-entangled photon pairs in a modified entangled quantum key distribution system, followed by all of the necessary classical post-processing including one-way error correction. We successfully exchange a 1,366\unit{bits} ROT string in $\sim$3\unit{min} and include a full security analysis under the noisy storage model, accounting for all experimental error rates and finite size effects. This demonstrates the feasibility of using today's quantum technologies to implement secure two-party protocols.
\end{abstract}

\maketitle


Cryptography, prior to the modern computing age, was synonymous with the protection of private communications using a variety of encryption techniques, such as a wax seal or substitution cypher. However, with the advent of modern digital computing and an increasingly internet-driven society, important new cryptographic challenges have arisen. People now wish to do business and interact with others they neither know nor trust. In the field of cryptography this is known as \emph{secure two-party computation}. Here we have two users, Alice and Bob, who wish to perform a computation on their private inputs in such a way that they obtain the correct output but without revealing any additional information about their inputs.

A particularly important and familiar example is the task of \emph{secure identification}, which we perform any time we use a bank's ATM to withdraw money. Here honest Alice (the bank) and honest Bob (the legitimate customer) share a password. When authenticating a new session, Alice checks to make sure she is really interacting with Bob by validating his password before dispensing any money. However, we do not want Bob to simply announce his password since a malicious Alice could steal his password and impersonate him in the future. What we require is a method for checking whether Bob's password is valid without revealing any additional information. While protocols for general two-party cryptographic tasks such as this one may be very involved, it is known that they can be built from a basic cryptographic building block called oblivious transfer (OT) \cite{Kil88}.

Many classical cryptography techniques currently in use have their security based on conjectured mathematical assumptions such as the hardness of finding the prime factors of a large number, assumptions which no longer hold once a sufficiently large quantum computer is built \cite{Sho94}. Alternatively, quantum cryptography offers means to accomplish cryptographic tasks which are provably secure using fewer assumptions that are ideally much more stringent than those employed classically. However, until now almost all of the experimental work has focused exclusively on quantum key distribution, yet there are many other cryptographic primitives \cite{Wie83,HBB99,GW07,BBBGST11} which can make use of quantum mechanics to augment their security. OT is a prime example.

While it has been shown that almost no cryptographic two-party primitives, save for QKD, are secure if a quantum channel is available and no further restrictions are placed on an adversary \cite{May97,LC97,Lo97}, two-party protocols are so pivotal to modern cryptography that we are required to explore scenarios that place realistic restrictions on an adversary which allow provable security to be restored in important real world settings. Moreover, if we can efficiently implement low-cost quantum cryptography while making it much harder and more expensive for adversaries to break it as compared to classical schemes, even if not provably secure without added assumptions, then we have still provided a benefit.

One can regain provable security for OT within the noisy-storage model \cite{WST08,KWW12} under the \emph{physical} assumption that an adversary does not possess a large \emph{reliable} quantum memory. This model most accurately captures the difficulty facing a potential adversary since most quantum memories currently suffer from one or more of the following problems: either the transfer of the photonic qubit into the physical system used for the memory is noisy; or the memory is unable to sufficiently maintain the integrity of the quantum information over time; or the memory suffers from an inability to perform consistently (ie. photons are often lost causing the memory to act as an erasure channel).

With current technology, the quantum information stored in a quantum memory is lost within a few milliseconds \cite{BRDRDSLLZP12,ZLYLG12,RLVBRD12}. While there have been recent demonstrations of systems with coherence times on the order of seconds and even minutes \cite{MKLJYBPHCMTCL12,SSSDRABPMT13}, high fidelity transfer of quantum information from another physical system into the memory system and its subsequent storage have not yet been demonstrated. Further, a reliable memory also requires fault tolerant error correction built into its architecture. Until these significant experimental challenges are met, security is assured under the noisy storage model. Even with the advent of practical quantum storage, for any storage size with a finite upper bound, security can still be achieved by simply increasing the number of qubits exchanged during the protocol. Moreover, the currently executed protocol holds secure even if a dishonest party obtains a better quantum memory in the future.

Because of its huge potential, securing OT with quantum means has recently received interest both from this experiment, as well as the recent 1-out-of-N OT experiment performed by Chan \emph{et al.} \cite{CLMST13}. We note however that, in contrast to our implementation, Chan \emph{et al.} achieves only a weak version of 1-out-of-N OT in which the attacker gains a significant amount of information. Additionally, continued theoretical work \cite{KWW12} has been used to propose a number of different experimentally feasible protocols that could be implemented using today's technology \cite{WCSL10,Sch10}. Notably, quantum bit commitment secure under the noisy storage model was recently shown by Ng \emph{et al.} \cite{NJMKW12}. And while it is indeed known that oblivious transfer can in principle be built from secure bit commitment and additional quantum communication \cite{Yao95}, the protocol for such a reduction is inefficient and has not been analyzed in a setting where errors are present.

In this work, we show the first experimental implementation of OT secured under the noisy-storage model using a modified entangled quantum key distribution system and all of the necessary classical post-processing algorithms including one-way error correction. During a $\sim$3\unit{min} quantum and classical exchange we generate 1,366\unit{bits} of secure ROT key accounting for all experimental error rates and finite size effects. Using a new min-entropy uncertainty relation to derive much higher OT rates we examine a number of trade-offs and analyze the secure parameter regimes. This demonstrates the feasibility of using today's quantum technologies to implement secure two-party protocols, most notably the building block necessary to construct a secure identification scheme \cite{DFSS07} to securely authenticate oneself at an ATM one day.

\section*{Results} \label{sec.Results}

\subsection*{The Oblivious Transfer Protocol} \label{sec.OTProtocol}

In a 1-2 OT protocol we have two parties, Alice and Bob. Alice holds two secret binary strings $\hat S_{0}, \hat S_{1} \in \{0,1\}^{l}$ of length $l$. Bob wishes to learn one of these two strings, and the string he decides to learn is given by his choice bit $C \in \{0,1\}$. The protocol is called oblivious transfer because of its security conditions: if Alice is honest, then in learning the string of his choice, $\hat S_{C}$, Bob should learn nothing about Alice's other string, $\hat S_{1-C}$; while if Bob is honest, Alice should not be able to discern which string Bob chose to learn, i.e. Alice should not learn C. Security is not required in the case where both parties are dishonest. The 1-2 refers to the fact that Bob learns one and only one of Alice's two secret strings.

In order to implement 1-2 OT we actually first implement 1-2 \emph{random} oblivious transfer (ROT) which is then converted into 1-2 OT with an additional step at the end. In the randomized protocol, rather than Alice choosing her two strings she instead receives the two secret binary strings $S_{0}$, $S_{1}$ $\in \{0,1\}^{l}$, chosen uniformly at random, as an output of the protocol. After the ROT protocol is complete, Alice can use these random strings as one-time pads to encrypt her original desired inputs $\hat{S}_{0}$ and $\hat{S}_{1}$ and send them both to Bob. Bob can then recover Alice's original intended string by using the $S_{C}$ he learned during the ROT protocol in order to decrypt his desired $\hat{S}_{C}$ thus completing a 1-2 OT protocol.

For the protocol to hold correct when both parties are honest, it is crucial that Alice and Bob have devices satisfying a minimum set of requirements, for example the loss and error rates have to be upper bounded. In the correctness proof, Alice and Bob's devices are assumed not to exceed a maximum loss and error rate. It is these upper bounded values that will be used in the secure ROT string formula (Eq.~\ref{eq.ROTKeyRate}) in order to determine the maximum allowable size for the generated ROT string. Whenever their devices fulfill these criteria, the protocol can be executed correctly. On the contrary, in the security proof, a dishonest party is assumed to be all powerful: in particular he/she can use perfect devices to eliminate losses and errors in order to use them as a cheating advantage instead. The only restriction on a cheating party in this model is that of the quantum memory accessible. We refer the reader to the formal security definition of this protocol in \cite{Sch10}.

\begin{figure*}[htbp]
    \centering
    \includegraphics[width=18cm]{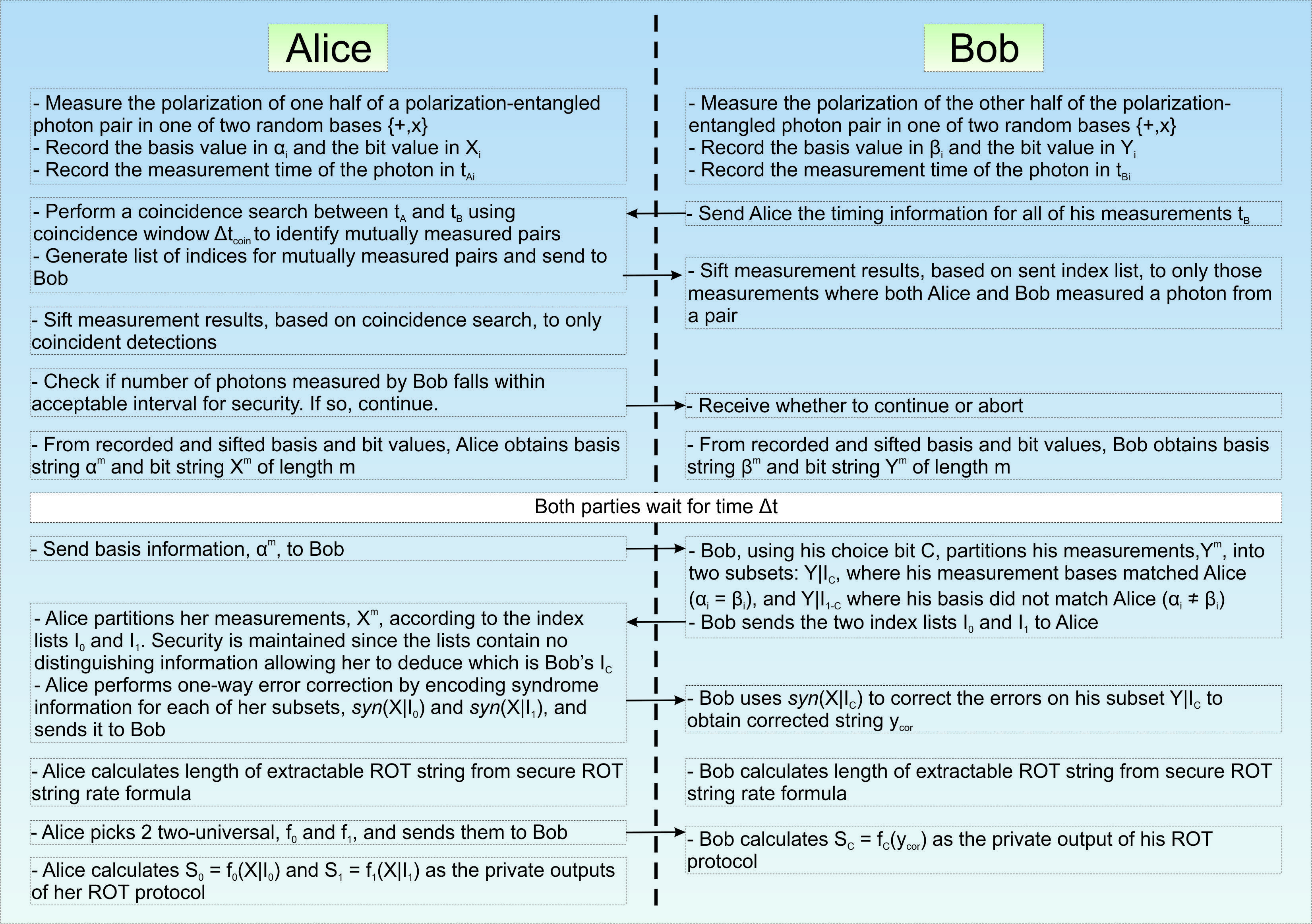}
    \caption{Flow chart of the 1-2 ROT protocol.}
    \label{fig.ROTProtocolFlowchart}
\end{figure*}

Fig. \ref{fig.ROTProtocolFlowchart} outlines our 1-2 ROT protocol \cite{Sch10,WCSL10}. A source of polarization-entangled photon pairs distributes one photon from each pair to Alice and Bob. In fact, Alice holds the entanglement source as it has been shown that this does not affect either correctness or security. Alice and Bob measure in one of the two usual BB84 bases, H/V (+) or $+45^{\circ}$/$-45^{\circ}$ ($\times$), using passive polarization detectors. For every photon they detect, they record their measurement basis as $\alpha_{i}$ and $\beta_{j}$, their bit value as $X_{i}$ and $Y_{j}$, and their measurement time as $t_{Ai}$ and $t_{Bj}$ respectively. Bob then sends his timing information to Alice so that she can perform a coincidence search allowing them to sift down to those measurement results where they both detected a photon from a pair. Alice then checks that the number of photons Bob detected falls within the secure interval allowing them to proceed. At this point Alice obtains basis and bit strings $\alpha^{m}$ and $X^{m}$, while Bob obtains basis and bit strings $\beta^{m}$ and $Y^{m}$, all of which are of length $m$.

Both parties now wait a time, $\Delta t$, long enough for any stored quantum information of a dishonest party to decohere. Note that the information Bob sends to Alice after this point does not provide her information about his choice bit $S_C$, provided that Bob's choice of basis is completely uniform. To ensure the uniformity of this choice, an honest Bob can perform symmetrization over all his detectors to make them all equally efficient \cite{NJMKW12}, so that Alice cannot make a better guess of his choice bit than a random guess. Hence, it is really only Alice who is waiting here to protect against a dishonest Bob who might have been storing some of his photons. Once the secure wait time is observed, Alice sends Bob her basis information, $\alpha^{m}$. This allows Bob to divide his bits into two subsets: $Y|\mathcal{I}_{C}$, where his measurement basis matched Alice's ($\alpha_{i} = \beta_{i}$) leading eventually to the shared bit string $S_{C}$; and $Y|\mathcal{I}_{1-C}$, where his measurement basis did not match Alice's ($\alpha_{i} \neq \beta_{i}$), which leads to the second bit string $S_{1-C}$ which Bob should know nothing about.

Next, Bob sends his index lists, $\mathcal{I}_{0}$ and $\mathcal{I}_{1}$, to Alice so that she can partition her data identically. Note that it is Bob's choice of labelling the subset where his measurement bases matched Alice's as $\mathcal{I}_{C}$ which allows him to eventually learn his desired $S_{C}$ from Alice since this is the subset in which error correction will succeed. However, this does not reveal $C$ to Alice since from her point of view she always receives an $\mathcal{I}_{0}$ and $\mathcal{I}_{1}$ from Bob and there is never any information in the partitioning which would allow her to deduce $C$.

Continuing, Alice performs one-way error correction by encoding syndrome information for each of her two subsets, $\mathrm{syn}(X|\mathcal{I}_{0})$ and $\mathrm{syn}(X|\mathcal{I}_{1})$, which she then sends to Bob. Bob is able to use the syndrome information for the subset where his measurement bases matched Alice's, $Y|\mathcal{I}_{C}$, to correct any errors in his string to obtain his error corrected string $y_{\mathrm{cor}}$. Note that Bob will only be able to correct his string if the devices and quantum channel are operating within their design parameters since this is what the one-way error correction code has been optimized for. If Bob is dishonest, then we do not need to worry about him being able to decode, the goal of the protocol is to ensure that \emph{honest} Bob can learn his desired string.

The last step is to perform privacy amplification \cite{CW79,Kra94}, which is a cryptographic technique that allows a situation of partial ignorance to be turned into a situation of (essentially) complete ignorance. For example, suppose Alice has a long string (here $X|\mathcal{I}_{1-C}$) which is difficult, but not impossible, for Bob to guess. That is, Bob's min-entropy about the string $X|\mathcal{I}_{1-C}$ is large. Applying a randomly chosen two-universal hash function $f_{1-C}$ to $X|\mathcal{I}_{1-C}$ allows Alice to obtain a (typically much shorter) string $S_{1-C} = f_{1-C}(X|\mathcal{I}_{1-C})$ which is essentially impossible for Bob to guess. That is, if the length of the short string is $2^{l}$, then Bob's guessing probability is very close to $1/2^{l}$ which means that Bob has learned nothing about the short string. The only cost to Alice for performing this is to reduce the size of her output strings somewhat according to Eq.~\ref{eq.ROTKeyRate}.

To perform privacy amplification, Alice and Bob use the secure ROT string rate formula (Eq.~\ref{eq.ROTKeyRate}) and the estimates of the error rate and loss due to their devices and the quantum channel to calculate the length of extractable ROT string which they can keep after privacy amplification. Alice applies the two-universal hash functions $f_{0}$ and $f_{1}$ to her two substrings $X|\mathcal{I}_{0}$ and $X|\mathcal{I}_{1}$ and obtains the shorter strings $S_{0} = f_{0}(X|\mathcal{I}_{0})$ and $S_{1} = f_{1}(X|\mathcal{I}_{1})$ as her private outputs, while Bob applies the appropriate two-universal hash function to his subset $y_{\mathrm{cor}}$ obtaining his desired $S_{C} = f_{C}(y_{\mathrm{cor}})$ as his private output. Privacy amplification is extremely important since Bob has potentially gained a significant amount of information about Alice's second string, which a secure protocol requires he know nothing about. This extra information has come from the syndrome information Alice sent for her second subset, $\mathrm{syn}(X|\mathcal{I}_{1-C})$, where their measurement bases did not match and from Bob's ability to store partial quantum information and attempt to cheat. In our proof, we show that $X|\mathcal{I}_{1-C}$ has a high min-entropy and hence the output $S_{1-C}$ is close to uniform after privacy amplification. Of course, dishonest Bob could use the error correction information to correct errors in his own storage device or he could open his device, remove all its imperfections, and then perform a partial attack to gain extra information without raising the error rate above the allowed limit. Regardless, by subtracting the entire length of the error correction information from the initial extractable ROT string Alice can extract a shorter string $S_{1-C}$ via privacy amplification, which looks uniform to Bob, ensuring the security of the protocol.

As the above protocol has just summarized, even though breaking the security of our protocol requires a large quantum memory with long storage times, neither quantum memory nor the ability to perform quantum computations are needed to actually run the protocol. Thus, as we shall see below, the technological requirements for honest parties are comparable to QKD and hence well within reach of current technology.

\subsection*{Experimental Parameters}\label{sec.ExpPara}

It is important to realize that the techniques used here to prove the security of ROT are fundamentally different than those used in proving the security of QKD. Contrary to QKD, there is no real-time parameter estimation which needs to be performed. Instead, the two parties estimate the parameters before performing the OT protocol. The only requirement is that it should be possible for honest parties to bring certain hardware to execute the protocol. Dishonest parties can have arbitrary devices and still security will be assured based on the following:
\begin{enumerate}
  \item Alice holds the source and can completely characterize it. Thus, if Alice is honest she can rely on the source following the estimated parameters. If she is dishonest, she is allowed to replace the source with an arbitrary quantum device (it could be a full quantum computer with arbitrary storage since the storage assumption is only needed to deal with a dishonest Bob).
  \item Honest Bob can always test his device himself without relying on Alice.
  \item The channel parameters (ie. the loss and error rate) can be estimated jointly by Alice and Bob. Depending on their estimate, Alice decides how much error correction information she needs to send. If Bob was dishonest and lied during the estimate then one of the following scenarios happens (both of which are secure):
      \begin{enumerate}
        \item Alice finds there is no $n$ such that secure OT can be performed in which case she either demands Bob get a better device or for the two of them to invest in a better channel.
        \item Alice tunes $n$ such that security can be obtained for their estimated parameters. Dishonest Bob could have lied and later eliminated his losses and errors for the OT exchange; however, this is already accounted for in the security analysis which assumes a dishonest Bob can have a perfect channel and perfect devices (his only limitation is the memory bound).
      \end{enumerate}
  \item If the parameters turn out to be different for the honest parties during the protocol (e.g. the errors turn out to be much higher) compared to what was estimated earlier, then the protocol may not succeed. Hence, once the parameters are estimated and fixed, then the honest parties need to use hardware satisfying these parameters if they want to perform the protocol correctly. However, security is not affected by the actual amount of losses and errors.
\end{enumerate}

To evaluate security we model our parametric down-conversion (PDC) entangled photon source in the standard way \cite{KB00} by measuring the mean photon pair number per pulse ($\mu$) which is directly related to the amplitude of the pump laser. Since it is a continuously pumped source we define our pulse length as the coherence time of our laser. Three other measured parameters are also required: the total transmission efficiency or transmittance, $\eta$, the intrinsic detection error rate of the system, $e_{\mathrm{det}}$, and the probability of a dark count in one of Alice's or Bob's detectors, $p_{\mathrm{dark}}$. The detection error is the probability that a photon sent by the source causes a click in the wrong detector, which can happen due to deficiencies or misalignments in Alice and Bob's equipment or due to processes in the quantum channel. For the dark count probability Alice and Bob take the value of their worst detector. Note that the parameters measured are necessary for allowing correctness of the protocol between two honest parties. In light of this, Alice and Bob perform a device and channel characterization and use these estimates in all subsequent security checks of the system. In the security analysis, a malicious party is assumed to have full power over their devices, such as eliminating all losses and noise in the communication while tricking the honest party to believe otherwise. Their values are summarized in the top of Table~\ref{tab.ExpParams}.

\begin{table}[htbp]
  \centering
  \begin{tabular}{cc}
      Experimental Parameters & Value \\
      \hline \hline
      $\mu$ & $(3.145 \pm 0.016) \times 10^{-5}$ \\
      $\eta$ & $0.0150 \pm 0.0001$ \\
      $e_{\mathrm{det}}$ & $0.0093 \pm 0.0002$ \\
      $p_{\mathrm{dark}}$ & $(1.50 \pm 0.04) \times 10^{-8}$ \\
      $n$ & $(8.00 \pm 0.06) \times 10^{7}$ \\
       & \\
      Adversary's Memory Limitations & Value \\
      \hline \hline
      $d$ & 2 \\
      $r$ & 0.75 \\
      $\nu$ & 0.002 \\
       & \\
      Security Parameters & Value \\
      \hline \hline
      $\varepsilon$ & $2.5 \times 10^{-7}$ \\
      $\varepsilon_{\mathrm{EC}}$ & $3.09 \times 10^{-3}$ \\
      $f$ & 1.491 \\
  \end{tabular}
  \caption{Experimental parameters. $\mu$ is the mean photon pair number per coherence time, $\eta$ is the total transmittance, $e_{\mathrm{det}}$ is the intrinsic error rate of the system, and $p_{\mathrm{dark}}$ is the probability of obtaining a dark count per coherence time; $d$ is the dimension of the assumed depolarizing channel (a qubit channel is assumed), $r$ is the probability that the memory (assumed to be a depolarizing channel) retains the state, and $\nu$ is the storage rate of the quantum memory ($\nu = 0$ means that no qubits can be stored, while $\nu = 1$ means all transmitted qubits can be stored); $\varepsilon$ is an error parameter used throughout the analysis, $\varepsilon_{\mathrm{EC}}$ is the error correction failure parameter, $f$ is the error correction efficiency, and $n$ is the total number (before losses) of entangled photon pairs exchanged. The correctness error for the protocol will be $2\varepsilon + \varepsilon_{\mathrm{EC}}$ and the security error will be $3\varepsilon$. The source is pumped with $7\unit{mW}$ of power, and Poissonian error bars and Gaussian error propagation have been used where appropriate.}
  \label{tab.ExpParams}
\end{table}

With these definitions we compute a number of probabilities needed for the security statements, which are \emph{conditioned} on the event that Alice observed a single click, namely: $p^{1}_{\mathrm{sent}}$ the probability that exactly one entangled photon pair is emitted from the source (all double pair emissions are assumed to give an adversary full information), $p^{h}_{\mathrm{B,noclick}}$ the probability an honest Bob did not detect a photon, and $p^{d}_{\mathrm{B,noclick}}$ the probability that a dishonest Bob did not detect a photon. Note that $p^{d}_{\mathrm{B,noclick}} \neq 0$ due to dark counts in Alice's detectors. Thus, even if a dishonest Bob removes all loss in the quantum channel and uses perfect detectors, there will still be pairs registered by Alice (she cannot differentiate between dark counts and valid detections of photons from entangled pairs) which a dishonest Bob misses. These probabilities are derived from our PDC model following Refs. \cite{Sch10,WCSL10} using the parameters given in Table \ref{tab.ExpParams} (more details can be found in the ``Methods: Experimental Security Parameters'' section).

\subsection*{Correctness}\label{sec.Correctness}

Whenever Alice and Bob are both honest, we desire that the protocol runs correctly even in the presence of experimental imperfections except with some failure probability. To quantify this probability, Alice and Bob can beforehand agree on some correctness error, $\varepsilon > 0$, which will be used to lower bound the failure probability of this protocol. This parameter can also be seen as a bound on the maximum allowable fluctuations observed during the protocol allowing one to form acceptable intervals around the expected values for $p^{1}_{\mathrm{sent}}$, $p^{h}_{\mathrm{B,noclick}}$, and $p^{d}_{\mathrm{B,noclick}}$ wherein correctness will hold. For example, the acceptable interval for $p_{\rm B, no click}$ is given by $[(1-p_{\rm B, no click}^{\rm h}-\zeta) n, (1-p_{\rm B, no click}^{\rm h}+\zeta) n]$ with
\begin{equation}
  \zeta = \sqrt{\frac{\ln \frac{1}{\varepsilon}}{2n}}
\end{equation}
given by invoking Hoeffding's inequality \cite{Uhl63}. According to Hoeffding's inequality, the number of detected rounds fall out of this interval with probability less than $2\varepsilon$. For our experiment, we chose a correctness error of $\varepsilon = 2.5 \times 10^{-7}$. Note that in our analysis there is a correctness error and a security error (discussed in the next section) which we represent by the same variable, $\varepsilon$, since they take the same value, but they could in general be different.

During the protocol Alice checks whether the number of rounds reported as lost by Bob lies outside the interval that would be expected based on the parameters estimated before starting the protocol. We emphasize that this test is not part of the parameter estimation, but rather ensures that the absolute number of rounds reported lost by Bob is limited. Intuitively, this is a critical step since it prevents a dishonest Bob, who has a perfect channel, from using the fact that he can report rounds as lost to discard some or all of the single photon rounds (for which he only potentially gains partial information) in favour of multi-photon rounds (where he gains full information). Thus, the step prevents a dishonest Bob from performing the equivalent of the well-known photon-number-splitting attack from QKD \cite{Hwa03} in our protocol. For more details, please refer to Ref.~\cite[Sec.~III]{WCSL10}.

Error correction is also necessary in order to correct errors due to the quantum channel, so that Bob can faithfully recover $S_{C}$. Error correction must be done with a one-way forward error correction protocol to maintain the security of the protocol. The error-correcting code is chosen such that Bob can decode faithfully except with a probability at most $\varepsilon_{\mathrm{EC}}$. Thus, our ROT protocol will succeed except with probability $2\varepsilon + \varepsilon_{\mathrm{EC}}$. For our error correction code, we have found an (empirically tested) upper bound to the error correction failure probability of $\varepsilon_{\mathrm{EC}} \leq 3.09 \times 10^{-3}$. Thus, the total correctness error is upper bounded by $2\varepsilon + \varepsilon_{\mathrm{EC}} \approx 0.0031$. Note that while the decoding error has a large effect on the total correctness error of the protocol, it does not affect the security error of the protocol. In particular, the security error can be much lower than the correctness error, which we will see in the next section.

\subsection*{Security} \label{sec.Security}

In this section, we show that the security error can be made small, i.e. the protocol holds secure against dishonest parties except with error $3\varepsilon=7.5\times 10^{-7}$. Note that in our analysis we first fix the security error, $\varepsilon$, and then derive the corresponding rate of OT. Thus, just like in QKD, one can reduce $\varepsilon$ at the expense of shortening the final OT string if a particular situation calls for a smaller security error.

To evaluate security for honest Alice, we need to prove that Bob does not know at least one of the extracted strings. This is done by quantifying Bob's knowledge about the string $X^m$ that Alice has. The min-entropy serves as a suitable quantification in such cases, and is defined as
\begin{equation}
H_{\rm min} (X|{\rm Bob}):=-\log p_{\rm guess} (X|{\rm Bob})
\end{equation}
where $p_{\rm guess} (X|{\rm Bob})$ is the probability of Bob guessing $X$ correctly, maximized over all possible measurements he can perform on his system. The min-entropy can also be understood operationally as the number of bits extractable from $X$, which appear random to Bob \cite{Ren08}. To allow for imperfect states we use the \textit{smoothed} min-entropy $H_{\rm min}^{\epsilon} (X|{\rm Bob})$ which is the min-entropy maximized over all joint states of $\tilde\rho_{X,{\rm Bob}}$ close (in terms of purified distance) to the original state $\rho_{X,{\rm Bob}}$, where $\rho_{X,{\rm Bob}}$ is the joint state of Alice's variable $X$ and Bob's entire system.

Note that a dishonest Bob's strategy, upon receiving information from Alice, can in general be as follows: he performs some arbitrary encoding upon the received state, and stores the state into some quantum memory (possibly retaining some additional classical information). After receiving Alice's basis information, he utilizes the stored qubits to perform strategic measurements again, gaining more classical information from his measurement outcome. Therefore, the main goal is to show that given all of Bob's information his guessing probability of $X^m$ is still low, i.e. Bob's min-entropy is lower bounded.

To do so, we employ the min-entropy uncertainty relation recently derived by Ng \emph{et al.} \cite{NBW12}, given by
\begin{equation}\label{eq.MinEntropy}
  {\rm H}^{\varepsilon /2}_{\mathrm{min}}(X^{m} | \alpha^{m} K) \geq m^1 \cdot c_{\mathrm{BB84}}
\end{equation}
where $\alpha^{m}$ is the vector of measurement bases of an honest Alice, $K$ represents some arbitrary classical side information, while $c_{\mathrm{BB84}}$ is a constant that depends solely on the measurements and arises due to bounding a family of {R\'enyi} entropies. The quantity $m^{1} = (p^{1}_{\mathrm{sent}} - p^{h}_{\mathrm{B,noclick}} + p^{d}_{\mathrm{B,noclick}} - 3 \zeta)n \approx 1.049 \times 10^{6}$ is a lower bound (including finite size effects) for the number of single photon pair rounds exchanged \cite{WCSL10}, except with probability $\varepsilon$. We use this to bound the minimum amount of uncertainty Bob has about Alice's overall string $X^{m}$. The value of $c_{\mathrm{BB84}}$ is given by
\begin{equation}\label{eq.MinEntropyC}
  c_{\mathrm{BB84}} = \max_{s \in (0,1]} \frac{1}{s}[1 + s - \log_{2}(1 + 2^{s})] - \frac{1}{s \cdot m_{1}}\log_{2} \Big( \frac{8}{\varepsilon^{2}} \Big).
\end{equation}

The bound on the min-entropy in Eq. \ref{eq.MinEntropy} also implies that, for security to hold at all we require
\begin{equation}\label{eq.PositiveMinEntropy}
  p^{1}_{\mathrm{sent}} - p^{h}_{\mathrm{B,noclick}} + p^{d}_{\mathrm{B,noclick}} -3\zeta > 0
\end{equation}
to generate a positive $m_{1}$ and by extension a positive min-entropy. This condition provides the regime of parameters, specifically with respect to $\mu$ and $\eta$, wherein secure ROT can be performed.

Subsequently, we further bound Bob's min-entropy conditioned on his quantum memory. ROT has been proven secure against adversaries whose memories satisfy the strong-converse property. This states that the success probability of decoding a randomly chosen n-bit string sent through the quantum memory decays exponentially for rates above the classical memory capacity. Recent theoretical results also show that this can be refined by linking security to the entanglement cost \cite{BFW12} and quantum capacity \cite{BCBW12} of the memory, instead of the classical capacity. To explicitly evaluate security, we follow Refs.~\cite{Sch10, WCSL10} and model the memory as a d-dimensional depolarizing channel, given by
\begin{equation}\label{eq.DepolChannel}
  \mathcal{N}_{r}(\rho) = r \rho + (1-r)\frac{\mathbb{1}}{d} \;\;\; \mathrm{for} \;\;\; 0 \leq r \leq 1
\end{equation}
which successfully transmits the state $\rho$ with probability $r$ and otherwise replaces it with the completely mixed state $\frac{\mathbb{1}}{d}$ with probability $1-r$, as the typical action of a quantum memory specializing to the case of qubits (ie. $d = 2$). A memory has one other pertinent parameter we need; namely, a storage rate, $\nu$, which represents the fraction of qubits which an adversarial Bob can store in memory. Intuitively, the noise ($1-r$) and storage rate ($\nu$) are related since increasing the amount of quantum error correction in the memory would decrease the noise at the cost of using more memory qubits for error correction thus decreasing the storage rate. However, for this analysis we leave them as two independent parameters. The choices for $r$ and $\nu$ in the middle of Table~\ref{tab.ExpParams} are made for an assumed future quantum storage device subjected to depolarizing noise that retains the state with probability $r = 0.75$. The storage rate $\nu = 0.002$ implies we assume that a dishonest party cannot store more than $\nu n \approx 1.600\times 10^5$ of all the $n = 8.00 \times 10^{7}$ qubits. Note that our experiment is quite secure as all current memories decohere after $\sim$1\unit{sec} and can only store a handful of photons.

There is a second condition necessary for security to hold given by
\begin{equation}\label{eq.CapacityBound}
  C_{\mathcal{N}} \cdot \nu n <  \frac{c_{\mathrm{BB84}} \cdot m_{1}}{2} - 1 - \log_{2}\Big( \frac{2}{\varepsilon} \Big)
\end{equation}
where $C_{\mathcal{N}}$ is the classical capacity of the memory ($\mathcal{N}$). This implies that the classical capacity of the total quantum memory (assumed to satisfy the strong converse property) should be strictly less than the total min-entropy of Bob's knowledge of $X^m$, including some finite size effects.

Finally, we can explore the ROT rate formula \cite{Sch10,NBW12} which tells Alice and Bob how many secure ROT bits they can keep from the privacy amplification operation. It is given by
\begin{equation}\label{eq.ROTKeyRate}
    l \leq \frac{1}{2} \nu \cdot \gamma^{\mathcal{N}} \bigg( \frac{R}{\nu} \bigg) \cdot n - f \cdot h(p_{\mathrm{err}}) \frac{m}{2} - \log_{2} \bigg( \frac{2}{\varepsilon} \bigg)
\end{equation}
where $\gamma^{\mathcal{N}}$ is the strong converse parameter of an adversary's memory ($\mathcal{N}$) which essentially gives the classical memory capacity, $R = \frac{1}{n}(c_{\mathrm{BB84}} \cdot m^{1} / 2 - 1- \log_{2}(2/ \varepsilon))$ is the rate at which a dishonest Bob would need to store quantum information, $m = (1 - p^{h}_{\mathrm{B, no click}})n \approx 1.128\times 10^6$ is the number of rounds where Alice and Bob both measured a photon from the pair, $f$ is the error correction efficiency relative to the Shannon limit, and $p_{\mathrm{err}}$ is the total probability of an error between Alice's and Bob's subset where their basis choices matched (including dark counts, their intrinsic detector error ($e_{\mathrm{det}}$), and any errors induced by the quantum channel). It is instructive to think of the three terms in Eq.~\ref{eq.ROTKeyRate} in the following way. The first term represents the amount of uncertainty Bob has over at least one of Alice's strings, regardless of the classical/quantum information he has. This is also the maximum ROT string that could be produced. From this one has to subtract the second term representing the information potentially leaked to an adversarial party during error correction, just as one would in QKD. This quantity is exactly the length of the error correction syndrome information which (as mentioned earlier) we must subtract from Bob's initial min-entropy to ensure security holds. Finally, the third term represents a safety margin guaranteeing the maximum security failure probability.

We conclude, that security holds for an honest Alice except with probability $3\varepsilon= 7.5\times 10^{-7}$, where the error comes from the finite size effects due to the number of single photon rounds, from smoothing of the min-entropy, and the error of privacy amplification. Note that the preceding security discussion used a simplified detector model and neglected vulnerabilities from a number of well known attacks in QKD \cite{GLLSKM11,LAMSM11,Mak09,QFLM07,MAS06} most of which target the detectors used.

\subsection*{Analysis of Secure Parameter Regimes}\label{sec.Security}

The ROT rate is sensitive to four parameters: the depolarizing noise, $r$; the storage rate, $\nu$; the intrinsic error rate of the system, $e_{\mathrm{det}}$; and the transmittance, $\eta$. Further, the secure range of each of these individual parameters depends on the values of the others, though we emphasize that the sensitivity is due in part to the fact that we tried to obtain the maximum length of OT string compatible with our $n$, $\varepsilon$, and storage assumptions. Similarly to QKD, one could make security less sensitive by being less demanding about the output length of the OT string. Nonetheless, we examine here the two most important parameters which determined the final ROT string length; namely, the transmittance and the error rate.

\begin{figure*}[htbp]
    \centering
    \begin{tabular}{cc}
      \includegraphics[width=8.5cm]{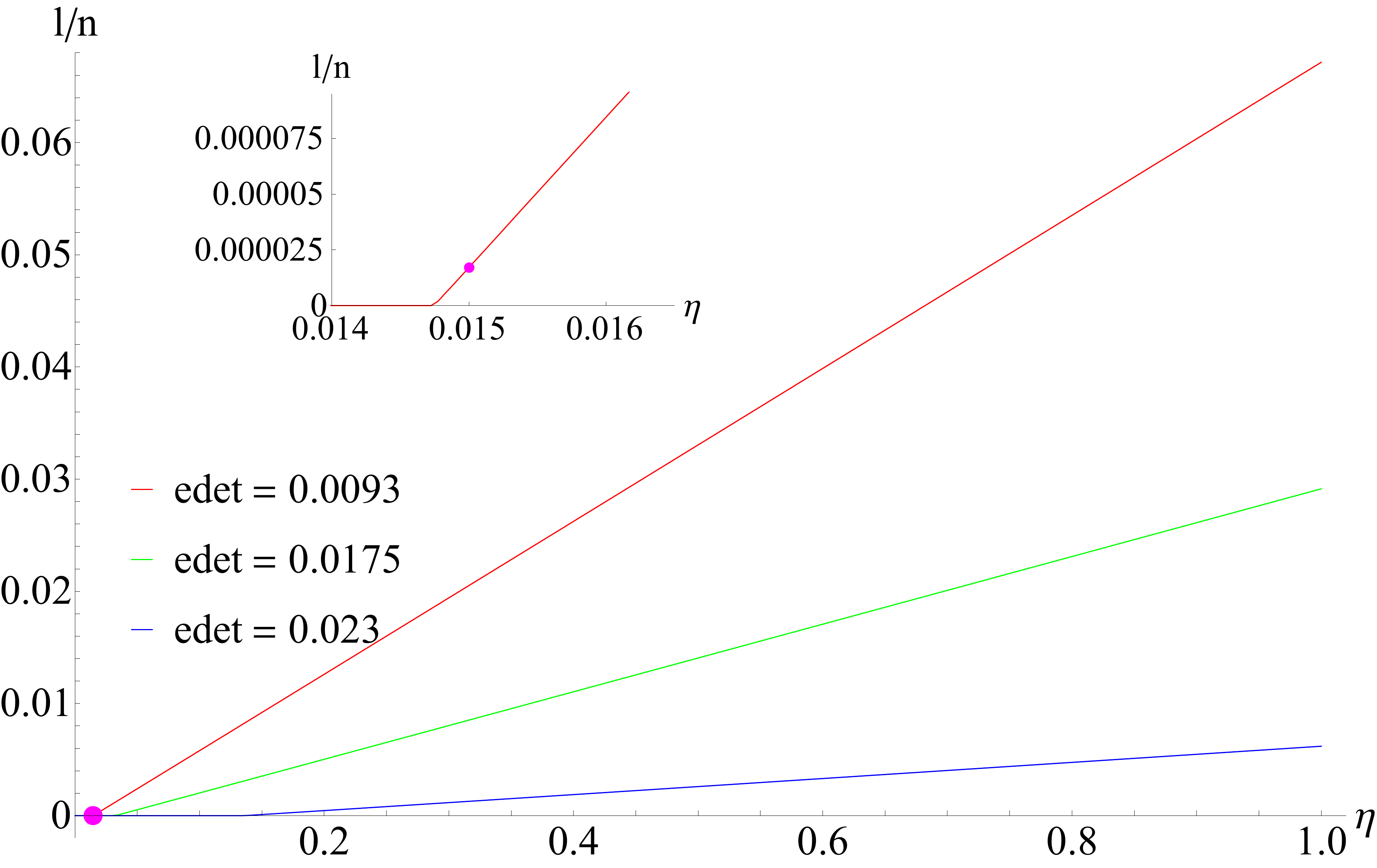} & \includegraphics[width=8.5cm]{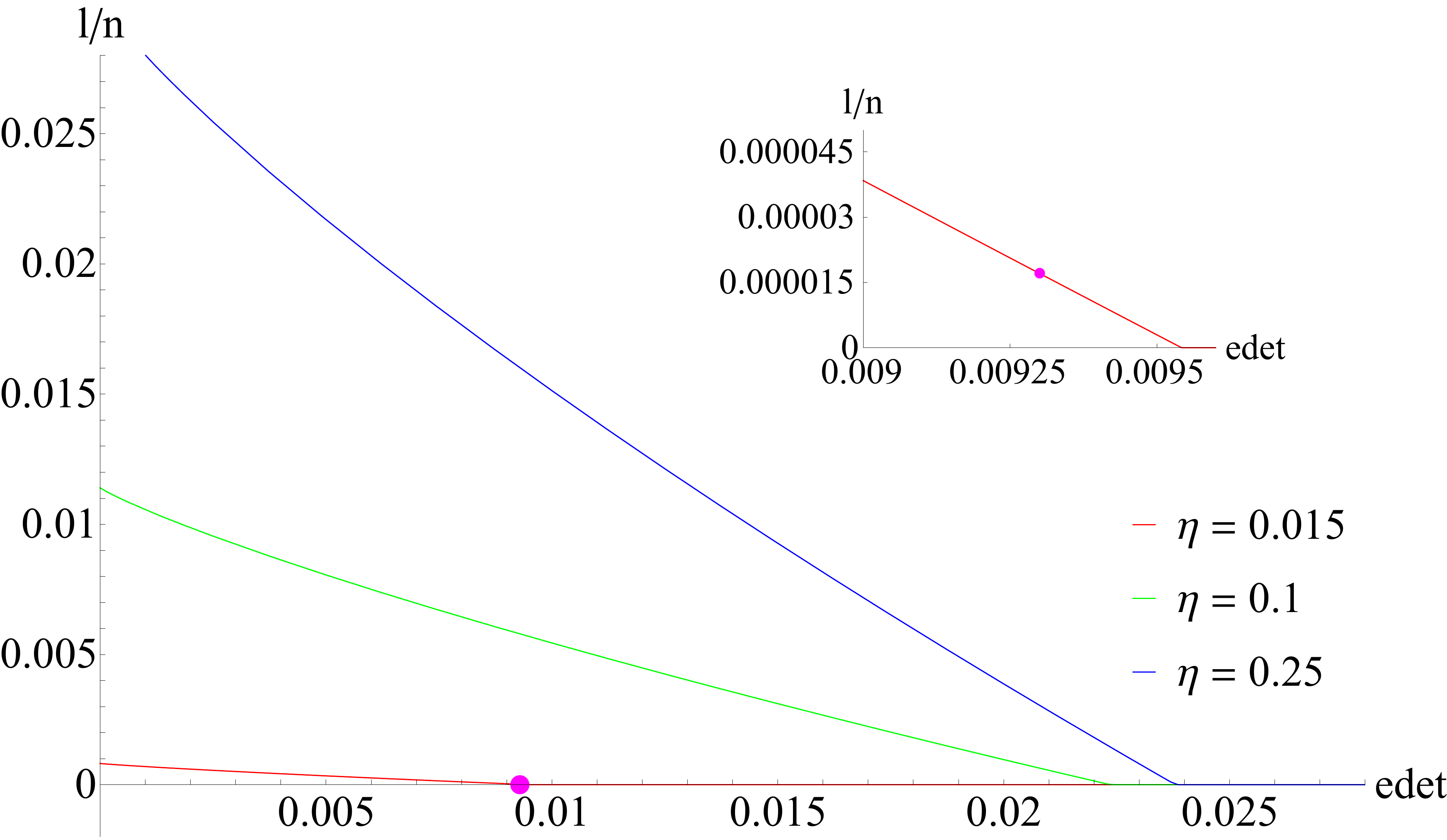} \\
    \end{tabular}
    \caption{Plot of the ROT rate, $l$, versus (left) the transmittance, $\eta$, for the $e_{\mathrm{det}}$ values of 0.0093 (red - this experiment), 0.0175 (green), and 0.023 (blue); and (right) the error rate, $e_{\mathrm{det}}$, for the $\eta$ values of 0.015 (red - this experiment), 0.1 (green), and 0.25 (blue). A magenta point is shown on the left graph representing the efficiency we experienced and the corresponding ROT string rate and on the right graph representing our error rate and the corresponding ROT string rate. Error bars are smaller than the data points.}
    \label{fig.lvsetaandedet}
\end{figure*}

Fig.~\ref{fig.lvsetaandedet} (left) shows the ROT rate versus the transmittance for error rates of 0.93\% (red - this experiment), 1.75\% (green), and 2.3\% (blue) from which we can see why loss is so crucial to the security of ROT. The secure ROT rate quickly drops as the transmittance decreases; indeed, for higher error rate values the ROT rate can quickly become negative even for relatively large transmittances, e.g. $\eta \sim 0.15$. Already with our table top experiment we experienced a total transmission efficiency of 1.5\%, which just allowed us to get a positive ROT string length. Our transmittance and the corresponding ROT rate are shown by the magenta point on the far left of this graph. The situation only gets worse when one thinks about using the scheme in a distributed quantum network where loss will be even higher.

Just as importantly, Fig.~\ref{fig.lvsetaandedet} (right) shows the ROT rate versus the detection error, $e_{\mathrm{det}}$, for transmittances of $\eta = 0.015$ (red - this experiment), $\eta = 0.1$ (green), and $\eta = 0.25$ (blue). From our system's transmittance of $\eta = 0.015$ we can see that in order to get a positive ROT rate we needed to observe an error rate of less than 0.954\% over the course of the experiment. Fortunately for this experiment, we managed to decrease the error rate to 0.93\%. Our error rate and the corresponding ROT rate are shown by the magenta point on this graph. While the constraint on the maximum allowable error rate does relax somewhat as the transmittance of the system increases, it never gets above a tolerable error rate of 2.38\% even at transmittances as high as 25\%. This represents an extremely experimentally challenging constraint, one that is much stronger than the typical safe QBER levels for QKD ($\sim 11\%$). This fact alone prevents most of the existing QKD systems in the world from being easily converted into a secure ROT implementation through a classical post-processing software update alone. Each of these two constraints can be individually relaxed at the cost of more stringent conditions on the other. However, our hope is that future theoretical work on the security of ROT in the noisy storage model can improve these bounds.

\subsection*{Experiment}

Before starting the experiment, all the pertinent parameters are estimated/selected and shown in Table~\ref{tab.ExpParams} in order for Alice and Bob to evaluate the length of the ROT strings, given by Eq.~\ref{eq.ROTKeyRate}, which they could subsequently extract from their data. Over the course of $\sim$50\unit{s}, Alice and Bob measured 1,347,412 coincident detection events. With the photon pairs distributed Alice verified that the number of Bob's reported measurements fell within the secure interval. After waiting for a minimum time $\Delta t = 1\unit{s}$ for any stored quantum information of a dishonest party to decohere, Alice sent her basis measurement information ($\alpha$) to Bob. Using his choice bit, $C = 0$, Bob partitioned his data into $Y|\mathcal{I}_{C=0}$ where his measurement basis matched Alice's (ie. $\alpha_{i} = \beta_{i}$), and $Y|\mathcal{I}_{1-C=1}$ where his measurement basis did not match Alice's (ie. $\alpha_{i} \neq \beta_{i}$), truncating both subsets to 600,000\unit{bits} (the block size needed for the error correction algorithm). He then sent his partitioning, $\mathcal{I}_{0}$ and $\mathcal{I}_{1}$, to Alice.

Alice partitioned her data accordingly, encoded the syndrome information for each subset, and sent it to Bob. With the system operating within the design parameter requirements, Bob was successfully able to error correct his subset of data, $Y|\mathcal{I}_{C=0}$, where his measurement basis matched Alice's in 2\unit{min} 14\unit{sec}. Using the the estimate of the error rate in Eq.~\ref{eq.ROTKeyRate}, Alice and Bob calculated the size of their secure final ROT string to be 1,366\unit{bits}. To complete the protocol, Alice then chose two 2-universal hash functions \cite{Kra94}, $f_{0}, f_{1} \in \mathcal{F}$, sent her choices to Bob, and calculated her private outputs $S_{0} = f_{0}(x|\mathcal{I}_{0})$ and $S_{1} = f_{1}(x|\mathcal{I}_{1})$ retaining the last 1,366\unit{bits} from the 2-universal hash operation in each case. Having chosen $C = 0$, Bob used $f_{0}$ to compute $S_{C} = f_{C}(y_{\mathrm{cor}})$ as his private output, obtaining the ROT string $S_{0}$ he desired from Alice.

Lastly, if Alice and Bob had wanted to remove the randomness from the protocol to implement OT with specific strings desired by Alice, she could have used $S_{0}$ and $S_{1}$ as one-time pads to encrypt her desired strings and sent the results to Bob. Using his $S_{0}$ he could then have decrypted Alice's last communication, recovering his desired string from Alice.

\section*{Discussion} \label{sec.Discussion}

It is important to notice that ROT is a fundamentally different cryptographic primitive than QKD and represents an important new tool at our disposal. Using the example of securing one's banking information at an ATM, it has long been suggested to use a small handheld QKD device to accomplish this \cite{DGHMR06}. However, if one were to employ ROT to build a secure identification scheme, as opposed to QKD, one's bank card number and PIN would never \emph{physically} be exchanged over any communication line. Rather, the ATM and the user would merely be evaluating the joint equality function, $f(x) = f(y)$, on each of their copies of the login information. This starkly illustrates the difference between ROT and QKD, as well as highlights ROT's potential in certain situations.

The use of the new min-entropy relation \cite{NBW12} was crucial to making our experimental implementation practical since it could be applied to much smaller block lengths while taking into account finite size effects. To put it in perspective, estimates using the original analysis from Ref. \cite{Sch10} suggested we required $n$ on the order of $10^{9}$ - $10^{10}$\unit{bits} requiring Alice and Bob to measure photon pairs for over 4\unit{hrs}. Any real-world secure identification protocol, such as at an ATM, making use of such an ROT implementation would obviously be entirely impractical. However, by employing the new min-entropy uncertainty relation we were able to generate positive OT lengths with an $n$ as low as $6.65 \times 10^{4}$. In fact, the limiting factor in our protocol became the minimum block size necessary for our one-way error correction code to succeed with high probability at the required efficiency.

Though the current work is secure for memories which follow the strong converse property, many important channels exhibit this property; for instance, the depolarizing channel assumed for this work. Moreover, while a few quantum memories have already been shown to work with high fidelity, they are all post-selected results. In other words, many photons are lost during their operation causing them to act as erasure channels. An adversary using one of these memories would find its action looking very close to a depolarizing channel since the best they could do is replace any lost photons with a random state. Moreover, recent theoretical results have shown this argument can be refined by linking the success probability of decoding to the entanglement cost \cite{BFW12} and quantum capacity \cite{BCBW12} of the memory, instead of the classical capacity.

There a number of other shortcomings that should be addressed in future work. As shown earlier, the ROT string length is constrained drastically by the transmittance, $\eta$. In order for ROT to be useful over longer distances, for instance over a future quantum communication network covering a city, the impact of loss on security has to be mitigated. One option could be to analyze the security proof for the case of an entangled photon source more carefully since a similar analysis for QKD has found that entangled systems can tolerate more loss than decoy-state prepare-and-send systems. Indeed, Wehner \emph{et al.} \cite{WCSL10} themselves point out they simplified their security proof for ROT in the case of an entangled source, giving all information in a double pair (and higher) emission to Bob. However, not only could the double pair emission rate be over-estimated in the assumed PDC model, but as has been pointed out in connection with QKD it is far from clear that double pair emissions give much, if anything, to an adversary. In fact the probability of double pair emission is one of the key quantities limiting the acceptable error rate, transmittance, and overall ROT rate; thus, there is likely much to be gained from a more detailed analysis. Another possibility to limit the impact of loss could be to find a secure version of Kocsis \emph{et al.}'s heralded noiseless amplification scheme \cite{KXRP13} which allowed both Alice and Bob to trust its operation.

Smaller issues to make ROT more practical include relaxing the tolerable error rate in order to use ROT in a long distance quantum network setting. Even at short distances with the new min-entropy relation \cite{NBW12} requiring us to detect as few as $6.65 \times 10^{4}$ entangled pairs in approximately $\sim$2.5\unit{s}, the extremely low error rate required makes it almost impossible to find an efficient error correction code at this block length.

Lastly, parameter estimation in the ROT protocol is fundamentally different than in QKD. Here parameters are estimated before the protocol begins and then the security statements are evaluated using the agreed upon experimental parameters. This does not present any security problems for the reasons outlined in Section ``Results: Experimental Parameters''; however, as a practical effect if the parameters are poorly estimated then honest parties will have a difficult time correctly executing the protocol. Heuristically it seems one would like to estimate the parameters for a long time such that the error bars on them are smaller than the intervals for Alice's security checks. This is not a problem since it only needs to be done once in advance and will not affect the number of rounds in the actual ROT protocol. In the case of devices (e.g. the source) which are not constant but have relatively slow fluctuations which would increase the error bars on the parameters during a long estimation, Alice and Bob could instead intersperse their ROT experiments with new parameter estimations. It is a very interesting open question for future work whether and how one could incorporate the error bars on the parameter estimates into the security proof.

In this work, we have shown the first experimental implementation of OT using a modified entangled quantum key distribution system and secured under the noisy-storage model. We performed an updated security analysis using a new min-entropy relation including finite statistics in order to show a drastic improvement in the secure ROT rate. We also examined the pertinent parameters which the ROT rate depends on, the most important being the transmittance and error rate. It will be very interesting to see whether future work on security proofs and experimental implementations can make the protocol even more practical in real-world scenarios.

\section*{Methods}

\subsection*{Experimental Implementation} \label{sec.ExperimentalImplementation}

The security proof for ROT guarantees security against adversaries with quantum memories and a quantum computer; however, the actual implementation of ROT for honest parties does not require these devices and is possible with today's technology \cite{WCSL10,Sch10}. Thus, we modified our existing entangled QKD system \cite{ECLW08,EHRLW10} to implement the first ROT protocol secured under the noisy storage model. The key difference to keep in mind is that while loss merely affects the overall key rate in a QKD system, in ROT loss is integral to the security of the scheme and if not properly bounded can prevent any secure ROT string from being generated.

\begin{figure*}[htbp]
    \centering
    \includegraphics[width=18cm]{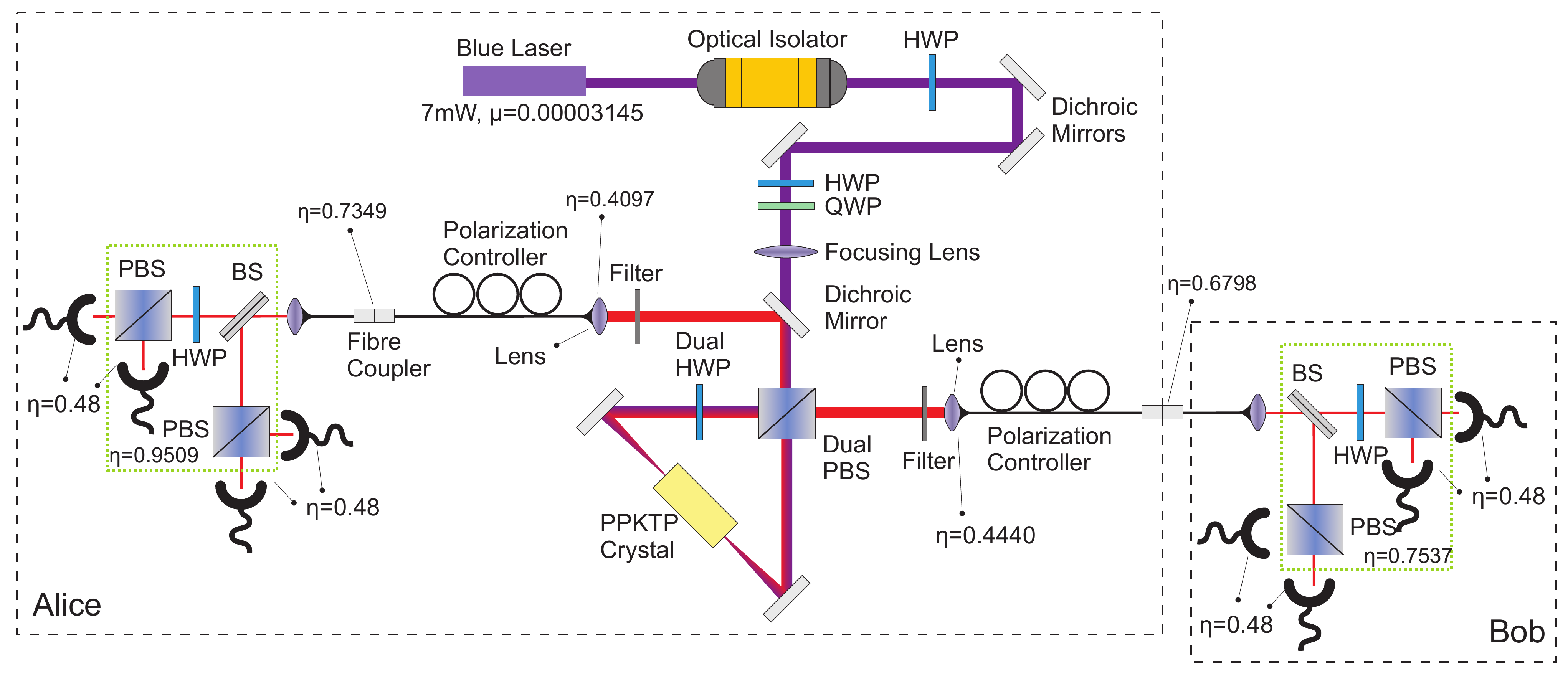}
    \caption{Schematic layout of the ROT experiment including all of the coupling losses experienced in the system. The source is pumped with $7\unit{mW}$ of power.}
    \label{fig.SystemLosses}
\end{figure*}

A schematic of the system is shown in Fig. \ref{fig.SystemLosses}. Entangled photon pairs are produced with a Sagnac interferometric entangled photon source \cite{KFW06,FHPJZ07,EHRLW10,PGW12} consisting of a periodically poled KTP (PPKTP) non-linear optical crystal produced for a collinear configuration placed in the middle of an interferometer loop. Entangled photons are produced by sending $45^{\circ}$ polarized light at 404\unit{nm} onto a dual wavelength polarizing beamsplitter (dual PBS) at the apex of the loop, thus bi-directionally pumping the PPKTP crystal. A dual wavelength half-waveplate (dual HWP) at $90^{\circ}$ on one side of the loop properly polarizes the pump laser so that the crystal produces down-converted polarization correlated photons at 808\unit{nm} in both directions around the loop. The dual HWP also rotates the polarization of the down-converted photons travelling clockwise around the loop by $90^{\circ}$ thus ensuring that when the photon pairs are split on the dual PBS their path information is erased and true polarization entangled photon pairs are generated. The first half of the entangled photon pairs are collected directly from one port of the PBS, while the second half of the pairs are collected via a dichroic mirror responsible for removing residual photons from the pump laser.

Since the ROT protocol is useful even over very short distances, such as to securely identify oneself to an ATM machine over $\sim 30 \unit{cm}$, Alice and Bob each measure their half of the entangled photon pairs while located next to the source connected to it with short single-mode optical fibres. The photons are measured with passive polarization detector boxes consisting of: a filter to reject background light, a 50/50 non-polarizing beamsplitter (BS) to perform the measurement basis choice, a PBS in the reflected arm of the BS to separate horizontally and vertically polarized photons, and a HWP and PBS in the transmitted arm of the BS to separate photons polarized at $+45^{\circ}$ and $-45^{\circ}$. Avalanche photodiode single photon detectors convert the photons into electronic pulses which are recorded with time-tagging hardware which saves the measured polarization and detection time with a precision of 156.25\unit{ps}. The data is transferred to Alice's and Bob's laptops where custom written software then performs the rest of the ROT protocol including entangled photon pair identification (based on the detection times), security checking, sifting, error correction, and 2-universal hashing of the final outputs.

While the majority of QKD systems, including ours, have used an interactive algorithm known as Cascade \cite{BS94} for error correction, we are not permitted to use it in ROT since the interaction would quickly reveal Bob's choice bit, C, to Alice. Instead, we chose to implement one-way forward error correction using low density parity check (LDPC) codes \cite{Gal62,MN97} updated for use in a QKD system \cite{Pea04,MCMHT09,Cha09}. LDPC codes are gaining popularity in a number of QKD experiments \cite{Pea04,MCMHT09} since they can relieve much of the classical communication bottleneck that interactive error correction algorithms cause. However, there is an important difference between one-way error correction in QKD compared to ROT; namely, in a QKD protocol error correction is permitted to fail on a block of key. In this case for QKD, Bob would let Alice know decoding failed on that block and they would have three options. They could either publicly reveal the block (in order to maintain an accurate estimate of the QBER) and then throw it away, they could try another one-way code optimized for a higher QBER, or they could revert to a two-way error correction procedure. We are not permitted to do any of these here as the error correction failure probability, $\varepsilon_{\mathrm{EC}}$, is the failure probability for running error correction on the \emph{entire} data set generated during the ROT protocol. A single failure of the error correction protocol requires Alice and Bob to start the protocol over.

In our LDPC-based error correction protocol, Alice and Bob share a large parity-check matrix, $H$, which is constructed using a modified version of Hu et al.'s Progressive-Edge Growth (PEG) software \cite{HEA05} with known optimal degree distribution profiles \cite{ELAB09,Mat11}. Alice computes $\mathrm{syn}(X|\mathcal{I}_{0})$ and $\mathrm{syn}(X|\mathcal{I}_{1})$ by treating her two bit strings as column vectors and applying $H$ to them. Using the syndromes sent to him from Alice, Bob then employs an iterative message passing (or belief propagation) decoding algorithm known as the sum-product algorithm to correct any errors between his bit string $Y|\mathcal{I}_{C}$ and Alice's $X|\mathcal{I}_{C}$ (where their basis choices matched) with high probability. Our sum-product LDPC decoder is written in C\# and is based on the Matlab code found in Ref. \cite{Cha09}. With a block size of $N = 600,000$ (matrix dimensions were $68000 \times 600000$) and average row weight of 40, we achieved an error correction efficiency of 1.491. Our particularly low source error rate, required to securely implement the protocol, made it very challenging to find an efficient code and necessitated the large block sizes. This in turn made decoding particularly time intensive with a single block taking on average 2\unit{min} 14\unit{sec}.

For the ROT experiment the source was pumped with 7\unit{mW} of power as a compromise between a high pair rate and minimal error rate. Alice and Bob exchanged timing information in real-time to sort down to coincidence events. These measurements formed their raw strings ($X$ and $Y$ respectively) combined with their record of measurement basis for each detection ($\alpha$ and $\beta$ respectively) and resulted in raw files 1.17\unit{MB} in size. It took Alice and Bob $\sim$50\unit{s} to measure 1,347,412 coincident detection events which meant that a total of $8.98 \times 10^{7}$ pairs (before losses) were distributed by the source. Due to slightly uneven detection probabilities and the inherent statistics of the source, Alice and Bob had to measure for slightly longer than the minimum time necessary to ensure that they measured 600,000 photon pairs with matching bases and 600,000 pairs with different bases. After error correction, Alice and Bob used an LFSR algorithm \cite{Kra94}, capable of operating on the \emph{entire} key at once, to perform privacy amplification. This is a key requirement to maintain security both in ROT and QKD, but one that is rarely, if ever, implemented in experimental and commercial QKD systems. All of the pertinent figures of merit from performing the ROT protocol are summarized in Table~\ref{tab.ExpResults}.

\begin{table}[htbp]
  \centering
  \begin{tabular}{cc}
      Figure of Merit & Value \\
      \hline \hline
      total \# of pairs created & $(8.00 \pm 0.06) \times 10^{7}$  \\
      \# of entangled pairs measured & 1,347,412 \\
      photon measurement time & $\sim$50\unit{s} \\
      $\Delta t$ & 1\unit{sec} \\
      size of $\hat{x}|\mathcal{I}_{0}$ & 600,000\unit{bits} \\
      size of $\hat{x}|\mathcal{I}_{1}$ & 600,000\unit{bits} \\
      error correction time & 2\unit{min} 14\unit{sec} \\
      size of ROT key & 1,366\unit{bits} \\
  \end{tabular}
  \caption{The pertinent figures of merit generated while performing the ROT protocol.}
  \label{tab.ExpResults}
\end{table}

\subsection*{Experimental Security Parameters} \label{sec.AppendixASecurityParams}

The security proofs for ROT \cite{Sch10,WCSL10} assumed that the states emitted by a PDC source can be written as \cite{KB00}
\begin{equation}\label{eq.AppPDCState}
  \ket{\Psi_{\mathrm{src}}}_{AB} = \sum^{\infty}_{n=0} \sqrt{p^{n}_{\mathrm{src}}} \ket{\Phi_{n}}_{AB}
\end{equation}
where the probability distribution $p^{n}_{src}$ is given by
\begin{equation}\label{eq.AppPDCProbDist}
  p^{n}_{\mathrm{src}} = \frac{(n+1)(\mu/2)^{n}}{(1+(\mu/2))^{n+2}},
\end{equation}
and the states $\ket{\Phi_{n}}_{AB}$ are given by
\begin{equation}\label{eq.AppPDCPhiStates}
  \ket{\Phi_{n}}_{AB} = \sum^{n}_{m=0} \frac{(-1)^{m}}{\sqrt{n+1}} \ket{n-m,m}_{A} \ket{m,n-m}_{B}.
\end{equation}
The states are written here in the basis $\ket{H,V}$ and $\mu/2$ is directly related to the amplitude of the pump laser resulting in a mean photon pair number per pulse of $\mu$. Using this assumption and a few other measured parameters, the authors of Refs \cite{Sch10, WCSL10} derived all of the other necessary quantities used in their proofs of security.

Following this prescription, the first thing we calculate is the parameter $\mu$ which is slightly more involved since we have a continuously pumped source which creates pairs at random times rather than the pulsed source assumed in the model. Using $7\unit{mW}$ of pump power, we measure Alice and Bob's single photon count rates per second, $N_{A}$ and $N_{B}$, as well as their coincident detection rate, $N_{coin}$, measured with a coincidence window of $\Delta t_{coin} = 3\unit{ns}$ (see Table~\ref{tab.AppSourceParams}). We then estimate Alice and Bob's total transmittances (including their source coupling, polarization analyzer, and detectors) using the formula $\eta_{A/B} = N_{coin}/N_{B/A}$ \cite{BGN00} which yields $\eta_{A} = 0.1374$ for Alice and $\eta_{B} = 0.1092$ for Bob. Using these numbers we back out the loss and estimate the total number of pairs produced at the crystal as $3,145,331\unit{pairs/s}$ with the formula $N = N_{A}/\eta_{A}$. For all count rates Poissonian error bars have been assumed, and Gaussian error propagation has been applied for the error bars on all derived quantities. Finally to calculate $\mu$, we define our pulse length as the coherence time of our laser since the security statements use $\mu$ with Eqs. \ref{eq.AppPDCState}, \ref{eq.AppPDCProbDist}, and \ref{eq.AppPDCPhiStates} to determine a bound on the possible double pair emission contributions which might expose additional information to a dishonest party. It is well known with PDC sources that pairs separated by more than a coherence time are independent and thus would not give a dishonest party additional information. It is only those pairs generated within the same coherence time which might pose a security risk. Thus, with a coherence length of $\Delta l_{c} = 3\unit{mm}$ for our laser (iWave-405-S laser specifications, Toptica Photonics \cite{Toptica}) and a corresponding coherence time of $\Delta t_{c} = 1.0 \times 10^{-11}\unit{s}$, we can calculate $\mu$ using the formula
\begin{equation}\label{eq.AppPDCMu}
  \mu = 3,145,331 \frac{\mathrm{pairs}}{\mathrm{s}} \times \Delta t_{c} = 3.145 \times 10^{-5}.
\end{equation}

\begin{table}[htbp]
  \centering
  \begin{tabular}{cc}
      Experimental Parameter & Value \\
      \hline \hline
      $N_{A}$ & 432,148 $\pm$ 657 \\
      $N_{B}$ & 343,470 $\pm$ 586 \\
      $N_{coin}$ & 47,197 $\pm$ 217 \\
      \hline \hline
      $\eta_{A}$ & $0.1374 \pm 0.0007$ \\
      $\eta_{B}$ & $0.1092 \pm 0.0005$ \\
      \hline \hline
      $N$ & 3,145,182 $\pm$ 16,148 \\
  \end{tabular}
  \caption{Source parameters for ROT where $N_{A}$ and $N_{B}$ are Alice and Bob's single photon count rates per second, $N_{coin}$ is their coincident detection rate per second, $\eta_{A}$ and $\eta_{B}$ are estimates for Alice and Bob's total transmittances, and $N$ is an estimate for the total number of pairs produced at the crystal per second. The source is pumped with $7\unit{mW}$ of power, and Poissonian error bars and Gaussian error propagation have been used.}
  \label{tab.AppSourceParams}
\end{table}

There are three other parameters that are necessary for the security statements: the total transmittance, $\eta$, which can be calculated from Fig.~\ref{fig.SystemLosses}; the intrinsic error rate (QBER) of the system, $e_{\mathrm{det}}$; and the probability of obtaining a dark count in one of Alice's or Bob's detectors, $p_{\mathrm{dark}}$. The dark count probability is defined similarly to $\mu$ (ie. by multiplying the dark count rate per second by the coherence time), and is taken as the value of their worst detector. Strictly speaking each detector will have a slightly different detection efficiency which will provide some partial information to the parties about the strings; however, this information can be removed by symmetrizing the losses of each user to their worst detector \cite{NJMKW12} so that they become independent of the basis choice. All of these values are summarized in the top half of Table~\ref{tab.ExpParams}.

The security statements (Eqs.~\ref{eq.MinEntropyC} -- \ref{eq.ROTKeyRate}) require the following important quantities: $p^{1}_{\mathrm{sent}}$, which is the probability only one photon pair is sent; $p^{h}_{\mathrm{B,noclick}}$, which is the probability that an honest Bob receives no click from a photon pair in his detector; and $p^{d}_{\mathrm{B,noclick}}$, which is the minimum probability that a dishonest Bob receives no click from a photon pair in his detector. All of these are derived from the PDC model given by Eq. \ref{eq.AppPDCState} with the experimental parameters given in Table \ref{tab.ExpParams}. For the derivations we refer the reader to Ref \cite{WCSL10}.


\begin{acknowledgments}
\textbf{Acknowledgements:}
Support for this work by NSERC, QuantumWorks, CIFAR, CFI, CIPI, ORF, ORDCF, ERA, and the Bell family fund is gratefully acknowledged. NN and SW are supported by MOE Tier 3 Grant MOE2012-T3-1-009. The authors would like to thank: Marcos Curty (Universida de Vigo) for many interesting discussions on the practical aspects of the security statements for real implementations of ROT; and Philip Chan (University of Calgary) for numerous explanations and discussions on implementing one-way forward error correction with LDPC codes as well as discussions on the security implications for using one-way codes.
\end{acknowledgments}


\bibliography{Paper8_Bibliography}
\bibliographystyle{naturemag}



\end{document}